# Influence of Electrode Thickness and Electrolyte Composition on SEI Formation and Cycle Stability in Binder-Free Hard Carbon Anodes for Na-Ion Batteries

Tien-Nguyen-Khoi Trinh [ł, ǂ,§], Adrian Beda [ł, ǂ,§], Samar Hajjar-Garreau [ł, ǂ], Cyril Vaulot [ł, ǂ], Camélia Matei Ghimbeu [ł, ǂ,§*]

[ł] Université de Haute-Alsace, Institut de Science des Matériaux de Mulhouse (IS2M), CNRS UMR 7361, 15 Rue Jean Starcky, Mulhouse 68100, France

[ǂ] Université de Strasbourg, Strasbourg 67081, France

[§] Réseau sur le Stockage Electrochimique de l'Energie (RS2E), CNRS FR3459, 33 Rue Saint Leu, Amiens Cedex 80039, France

*E-mail: camelia.ghimbeu@uha.fr.

# Table of Contents

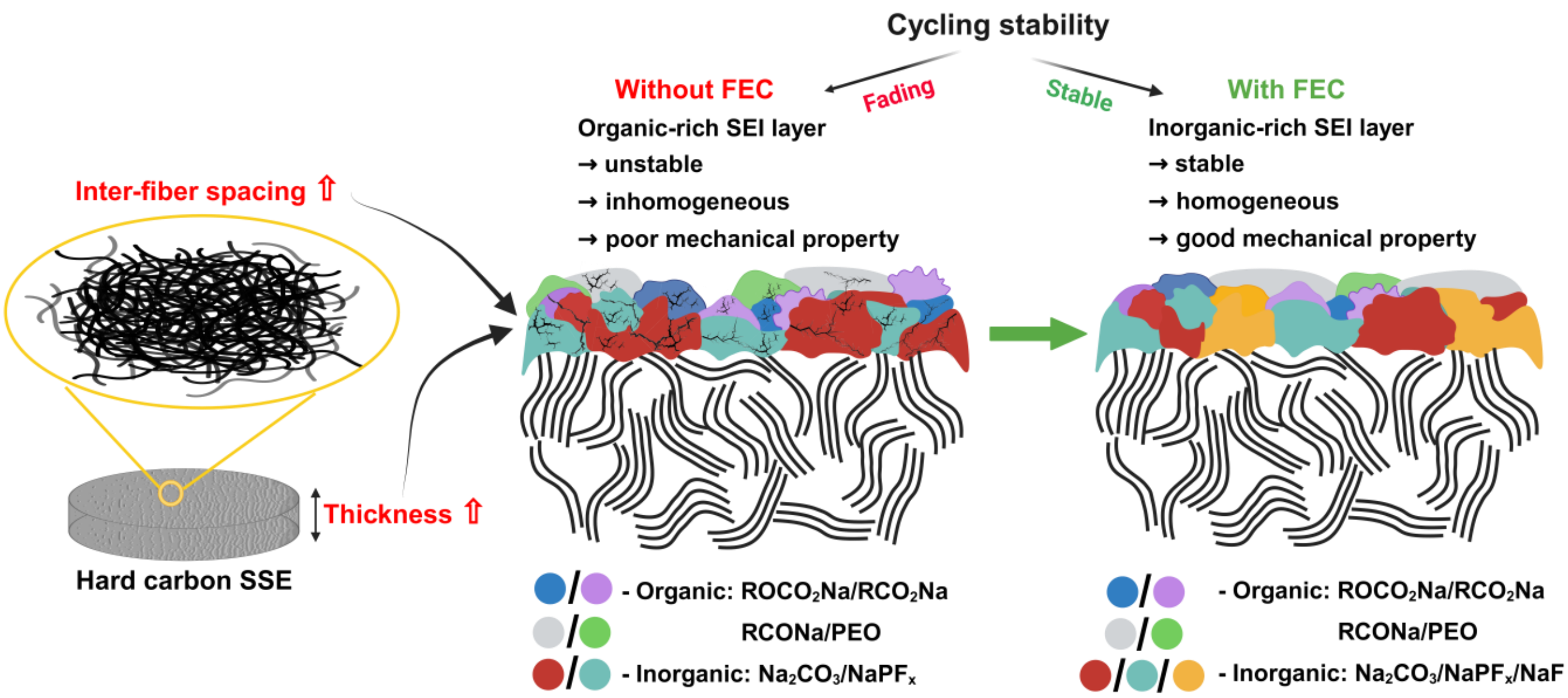

## Abstract

Hard carbon self-supported electrodes (HC SSEs) have gained significant interest as an alternative to classical binder-based anode materials for sodium-ion batteries (SIBs), eliminating the need for binders, solvents, conductive carbon, and current collectors. In this work, we investigate the impact of HC thickness and macroporosity, as well as the presence/absence of an electrolyte additive, on the formation of the solid electrolyte interphase (SEI) and electrochemical performance. Three cellulose-based filter paper (FP) precursors with different thicknesses were used to prepare HC SSEs through a pyrolysis process at 1500 °C, and then these were compared with an HC SSE obtained from commercial carbon fabric. The results highlight the importance of appropriate structural, morphological, porosity, and surface chemical properties of the materials to achieve high initial Coulombic efficiency (up to 93%) and high reversible capacity (up to 328 mAh $g^{-1}$ at 37.2 mA $g^{-1}$). However, the cycle stability was greatly impacted by the electrode thickness/macroporosity and the electrolyte additive. In $NaPF_6$ without fluoroethylene carbonate (FEC) additive, the maximum number of cycles (50) was obtained for the thinnest electrode (79 μm), while rapid fading (after ≈1–5 cycles) was observed for electrode thicknesses >185 μm. By using the FEC additive, the fading is mitigated and cycling stability (100 cycles) is reached for all the electrodes, except the commercial one (5 cycles), due to its highest thickness (1042 μm). The improvement in stability could be explained by the buildup of a thin homogeneous inorganic-rich SEI containing more NaF in the presence of FEC, which is more stable, protective, and electronically conductive.

# 1. Introduction

The transition to a carbon-neutral society requires significant advances in electrical energy storage technology, especially alkali-ion batteries, to replace fossil fuels or nuclear energy sources with green energy from renewable sources (e.g., wind, tidal, solar, etc.). Over the past few decades, lithium-ion batteries (LIBs) have evolved significantly and are widely used in our daily lives. However, due to safety concerns and geopolitical issues related to the scarce availability of metal resources (lithium, cobalt, and nickel), alternative technologies must be developed. Among them, sodium-ion batteries (SIBs) represent promising candidates, thanks to the abundant sodium resources, their even distribution worldwide, their lower cost, and their similar physicochemical properties to lithium.[1] Additionally, the use of cheaper and lighter Al current collectors instead of Cu, at both electrodes, avoids the formation of alloys below 0.1 V vs $Na^+/Na$;[2] therefore, SIBs can be discharged to near 0 V (zero energy) without causing damage, making them safer than LIBs.[3] Recent developments in electrode materials have enabled SIBs to meet the sustainability and cost-effective operating requirements for stationary energy storage systems.[4] However, challenges remain in the commercialization of this technology and include the development of low-cost electrode materials (anode/cathode) and safer electrolytes, as well as the enhancement of energy density.

Among the negative electrodes tested for SIBs (carbon-based materials, Ti-, Sn-based materials, etc.),[5] hard carbon (HC) attracts special attention due to its advantages: low cost, availability, and theoretical capacity close to graphite (in LIBs).[6] HC is defined as a disordered material with a complex microstructure, including randomly oriented amorphous and pseudographitic domains resulting in open/closed pores, described by Dahn et al.[7] as the "falling card model", in addition to heteroatoms, functional groups, and defects. Generally, HC is synthesized through a pyrolysis process, using different available resources, with renewable biomass being the most employed one. Regarding electrode preparation, HC powder is usually mixed with a conductive agent and a polymer binder in the presence of a solvent, to provide mechanical stability and good adhesion to the current collector (CC). However, the most used binder (polyvinylidene fluoride, PVDF) is difficult to recycle and requires a toxic and volatile solvent (N-Methyl-2-pyrrolidone, NMP) for dissolution, leading to environmental concerns.[8] Therefore, studies using green binders formulated in water, are becoming more popular. Among them, Carboxymethyl Cellulose (CMC) is often used since it presents advantages such as lower cost, biodegradability, better adhesion between active materials and CC (related to strong bonding by hydroxyl and carboxyl groups), and better electrolyte decomposition control.[9] Nevertheless, the binder presence in electrode formulation can also sometimes cause limitations related to increased electrode weight (dead mass due to its insulating properties), compatibility with the active material and reduced performance.[10]

In this respect, HC binder-free self-supported electrodes (HC SSEs) represent an appealing alternative to eliminate the use of binders (electrically insulating in nature), solvents (toxic), carbon black additive (irreversible reactions with the electrolyte), and current collectors (expensive), thus reducing the battery weight and cost, and enhancing the performance.[10] The manufacturing process of HC SSEs is often simpler and greener, and facilitates the recycling process. With the entire electrode component being the active material, the electronic conductivity as well as the energy and power density performances can be enhanced.[11] Thanks to these advantages, studies on HC SSEs for SIBs have gradually gained more attention in recent years (see **Table S1, Supporting Information, SI**). Ren and coworkers[12] synthesized HC SSEs from synthetic polymers (polyacrylonitrile—PAN combined with polyethylene glycol—PEG) pyrolyzed at 1100 °C. Their MCCF-5 electrode, using 1 M $NaClO_4$ in ethylene carbonate:dimethyl carbonate (EC:DMC, 1:1) electrolyte, delivered an initial reversible capacity (RC) of up to 366.5 mAh $g^{-1}$ at 25 mA $g^{-1}$, and good stability (90%), up to 300 cycles, at 100 mA $g^{-1}$. However, the achieved initial Coulombic efficiency (iCE) was rather low (73.7%), which may be related to the high specific surface area (SSA) of the sample (SSA $CO_2$ ≈ 651 $m^2$ $g^{-1}$). This increased the electrolyte decomposition reaction at the electrode interface, thereby increasing the irreversible capacity.[13] Huang et al.[14] reported the synthesis of composite free-standing carbon electrodes containing glucose-based HC and graphitic carbon via Ni foam (HC/NF). First, the precleaned Ni foam was immersed in a glucose solution, subjected to hydrothermal treatment at 180 °C for 18 h, then washed, dried, and carbonized at 800 °C for 1 h under Ar atmosphere. The HC/NF-2 electrode, tested with 1 M $NaClO_4$ in EC:DMC (1:1) with 5% fluoroethylene carbonate (FEC) electrolyte, was capable of delivering a good RC of 316.7 mAh $g^{-1}$ at 0.1 A $g^{-1}$ in the first cycle, despite a very low iCE (38.3%). The high RC achieved by this electrode is explained by the graphitic carbon-rich structure on the HC, which significantly enhances its conductivity. Meanwhile, the excessive solid electrolyte interphase (SEI) layer formation caused by irreversible reactions of the electrolyte with HC surface functional groups, confirmed from their cyclic voltammetry (CV) results, leads to very low iCE values. Nevertheless, its stability reached up to 1000 cycles at 1.0 A $g^{-1}$, while maintaining a specific capacity of 166.2 mAh $g^{-1}$ with CE > 99%. Gao et al.[15] used a different approach and produced SSEs by pyrolysis of spent face masks. The electrodes could deliver an iCE of 86% and an RC of 280 mAh $g^{-1}$ at 20 mA $g^{-1}$, with stability up to 300 cycles at 200 mA $g^{-1}$, when using 1 M $NaPF_6$ in Diglyme as electrolyte.

Most of the reported studies on HC SSEs show that effectively limiting the initial irreversible capacity while providing a high reversible capacity and stable cyclability remains quite difficult to achieve, with only a few cases being able to simultaneously deliver an iCE >90% with RC >300 mAh $g^{-1}$ and good cycling stability (such as H-THC or HCP, etc.), as noted in **Table S1,**

**SI**.[16,17] Such good performance could be obtained by HC SSEs that possess suitable interlayer spacing and disorder degree/defects, low oxygen content and specific surface area along with hierarchical pore structure (micro/meso/macropores) in some cases.[16,18,19] In addition, the choice of electrolyte (ether-based or ester-based) also contributes to a significant change in the performance of HC SSEs (iCE, reversible capacity, and cycling stability), as reported in the literature.[17,19] In another work, MFCP–1100 electrode performance was compared in ether and ester-based electrolytes and showed higher iCE and reversible capacity (95 vs 84.3% and 332.3 vs 297 mAh $g^{-1}$), as well as better cycling stability (500 cycles at 1 A $g^{-1}$ vs 300 cycles at 0.2 A $g^{-1}$) with the ether-based electrolyte (**Table S1, SI**).[20] The H-THC, HCP and MFCP–1100 carbons mentioned above delivered high performance, but in diglyme-based electrolytes.[16,17,20] Such electrolytes are not stable at high voltages when used to build full cells, and specific additives are required.[21] Besides, to our knowledge, no work showing the influence of the additive presence (such as FEC) in ester-based electrolytes on HC SSE was reported. Moreover, the impact of SSE thickness and macroporosity was not addressed either. Thicker electrodes with higher mass loading are of high importance for real-battery application, since they can improve the areal capacity.[22] However, they might adversely affect ion and electron transport, inducing larger polarization and capacity fading; therefore, a compromise must be found.[23] Besides, the study on the interaction between anode materials and electrolytes, which leads to the formation of SEI, is also essential to improve the performance and lifespan of SIBs. Nevertheless, studies on the passivation layer properties of HC SSEs in SIBs are rather rare. When analyzing the SEI layer of a conventional HC electrode, the presence of binder, conductive carbon, and the current collector will contribute additional signals to the data acquisition and complicate its analysis. For example, the C–O functional groups of CMC are difficult to distinguish from the C–O groups of the SEI layer by X-ray photoelectron spectroscopy (XPS) technique.[24] Alternatively, in temperature-programmed desorption mass spectrometry (TPD-MS) analysis, PVDF degrades, complicating the identification of gases released from carbon and SEI, such as $CO_x$, fluorinated compounds, etc.[25] Therefore, an SSE is considered as an ideal material for understanding the intrinsic interaction between HC and electrolyte at the interface level.

In this work, we propose a simple approach for HC SSE fabrication from cellulose filter paper (FP) to be used as model materials to understand the effect of electrode thickness and macroporosity on SEI formation and performance vs $Na^+$/Na, in the presence/absence of FEC additive. Three types of cellulose-based FPs with different thicknesses were used and compared with a commercial HC fabric. The SSE thickness and the additive prove to be crucial to achieve both high performance (iCE and RC) and cycling stability. In particular, we have attempted to identify the responsible parameters for stability behavior and understand the beneficial role of FEC by in-depth analysis of the SEI layer by XPS. The results highlight the importance of electrode thickness/macroporosity (interfiber spacing) on capacity fading, while the FEC additive helps in building a robust SEI that maintains cycling stability.

# 2. Experimental Part

### 2.1 Material Synthesis

Three types of cellulose-based filter paper (FP) with different thicknesses and properties were purchased from Rotilabo: Type 15A, and Whatman: grade 3 and grade 2589A. Kynol carbon fabric CC-509 (phenolic novolac resin-based) was graciously provided by Kynol Europa GmbH, Germany, and shown in **Scheme S1, SI**. The thickness characteristics of these four precursors, along with the chemical properties of commercial Kynol carbon fabric, are detailed in **Tables S2 and S3, SI**. The synthesis process previously reported by our group was optimized (use of different precursors, pyrolysis temperature, and duration) with the aim of simplifying it (removal of impregnation/thermopolymerization steps) and improving the performance.[10] The FPs were first punched to a diameter of 1.6 cm, then a two-step pyrolysis was performed (illustrated in **Scheme S1, SI**): 1) at 900 °C for 1 h under Ar (15 L $h^{-1}$), where the sample was fixed between two alumina plates to avoid bending, followed by natural cooling; 2) at 1500 °C for 3 h, under Ar (same flow), with the carbon electrodes obtained being placed between glassy carbon plates to avoid any interactions that could be caused by alumina. Kynol carbon fabric was punched to a diameter of 1.2 cm and then directly heat-treated at 1500 °C for 3 h. With each initial thermal decomposition process at 900 °C, six samples can be prepared per batch, while for the supplementary thermal decomposition process at 1500 °C, a maximum of 14 samples can be obtained per batch, with good reproducibility (as illustrated in **Figure S1, SI**). The obtained HC SSEs with a diameter of ≈1.1 cm were denoted as 15A, G3, GA, and Kynol, respectively. Images of these four samples are provided in **Figure S2, SI**. The as-obtained materials were directly used as working electrodes in Na-ion half-cells.

### 2.2 Material Characterization

The morphology of HC SSEs was observed by scanning electron microscopy (SEM, FEI Quanta 400) using a high-resolution low-vacuum field emission gun (FEG). The fiber diameter size distribution of the materials was determined with ImageJ software by counting ≈600 fibers from several in-plane SEM images for each sample. In addition, the thicknesses of the HC SSEs were determined through cross-sectional SEM images. The chemical composition of the precursor Kynol Carbon fabric CC-509 and the HC SSEs was obtained using an Oxford INCA X-Act energy-dispersive spectrometer (EDS).

Regarding the textural properties, mercury intrusion porosimetry (MIP) analysis was performed, using a Micromeritics Autopore IV-9500, to provide information on material’s meso- and macroporosity. The sample preparation and MIP analysis are described in more detail in the **Supporting Information**. Besides, the gas adsorption method at standard temperature and pressure (STP) was used to study micro- and mesoporosity of the HC SSEs through Micromeritics ASAP 2420 and ASAP 2020 instruments, with $N_2$ (77 K) and $CO_2$ (273 K) adsorbates, as detailed in the **SI**. High-resolution transmission electron microscopy (HR-TEM), X-ray diffraction (XRD), and Raman spectroscopy were used to evaluate the structure of HC SSEs. The JEOL ARM-200F instrument, working at 200 kV, provided HR-TEM images at a resolution down to 20 nm. XRD diffractograms were collected, in the angular region of 10° ≤ 2θ ≤ 90°, by a Bruker D8 Advance~~d~~diffractometer (Cu $K\alpha_{1,2}$ λ = 1.5406 Å) adopting a flat-plate Bragg–Brentano θ-θ geometry. The graphene interlayer spacing ($d_{002}$) was calculated by using the Bragg's law on the peak of the (002) planes, while the crystallite height ($L_c$) and the crystallite length ($L_a$) values, corresponding to the thickness and width of the coherent pseudographitic domains, were determined by applying the Scherrer equation. The number of graphene layers ($N$) was calculated based on $L_c$ and $d_{002}$ as described elsewhere.[6]

A LabRAM 300 (Horiba Jobin-Yvon) spectrometer was used to measure nine Raman spectra in different areas for each sample (mapping mode). The instrument was equipped with a continuous-wave (CW) diode-pumped solid-state laser (wavelength = 532 nm), and the following analysis parameters were set: objective lens (50x), filter (D1), and acquisition time (60 s). The average spectrum was fitted in Origin software with a combination of a Lorentz function for *D*-band (disordered domain) and a Gaussian function for *G*-band (graphite domain), using the procedure of Sadezky et al.,[26] to which we applied some modifications. More specifically, the *D*-band was fitted with a three-peak model (*D*1, *D*3, and *D*4).[27] After the deconvolution of these peaks, their area values were used to determine the area ratio ($I_{D1}/I_G$) of the *D*1 and *G* peaks to assess the disorder degree of the HC materials.

To identify and quantify the oxygen-based functional groups ($CO_x$), as well as the active surface area (ASA), temperature-programmed desorption (TPD) was performed in a vacuum system equipped with a mass spectrometer (MS).[28] Details of the MS calibration and TPD-MS/ASA analysis are provided in the **Supporting Information**.

### 2.3 Electrochemical Performance Evaluation

Coin cells (CR2032-type) were used for electrochemical testing. All HC SSEs, with a diameter of 1.1 cm and a mass loading of ≈2–6 mg $cm^{-2}$ for the electrodes (except Kynol, which is much higher at ≈30 mg $cm^{-2}$), were first dried under vacuum at 120 °C for 12 h in a Büchi oven to remove any traces of moisture. The half-cells were then assembled in an argon-filled glovebox, with the following components: sodium metal (Sigma-Aldrich, 99.9% purity, d = 1.2 cm) as the counter/reference electrode; two glass fibers (Whatman GF/C, d = 1.6 cm) as the separator; and HC SSEs as the working electrode. Two different electrolytes were used to study the electrochemical performance of the materials: (i) 1 M $NaPF_6$ dissolved in a mixture of EC:DMC with equal volume fractions and (ii) the same electrolyte solution, but containing 2% of FEC additive. The BioLogic BCS-805 battery cycler was used to measure the performance in a voltage window of 0.005–2.1 V (versus $Na^+/Na$), with the applied current density (C-rate) considered to be 1C = 372 mA $g^{-1}$. The electrochemical analyses were carried out as follows: (i) galvanostatic step (constant current at 0.1C) with voltage down to the lower limit of 0.005 V; (ii) potentiostatic step (constant voltage at 0.005 V) with the current density reduced to 0.01C; and (iii) the current was reversed, and a galvanostatic step at 0.1C was performed with voltage increased to the upper limit of 2.1 V. The initial rate of the process was maintained at 0.1C for the first five discharge/charge cycles, then it was changed to 1C. All experiments were conducted at room temperature, and the potentials were compared with Na metal (vs $Na^+/Na$). Electrochemical measurements of each sample were performed with at least two cells to ensure reproducibility, with an error of less than 5%. Capacity retention (CR) of HC SSEs was calculated from the ratio between the reversible capacity of the 100th cycle and the 7th cycle at 1C. The reversible capacity decay rate per cycle is calculated and plotted using the following formula:

$$\text{RC decay per cycle (\%)} = \frac{RC_{n-1} - RC_n}{RC_{n-1}} \times 100 \tag{1}$$

The current density regime test was carried out under symmetrical conditions by varying the C-rate of two galvanostatic steps as follows: 0.1C, 1C, 3C, 5C, 10C, and back to 1C (5 cycles at each C-rate). A BioLogic VSP-300 multichannel potentiostat was used to measure the electrochemical impedance of the HC SSEs.

### 2.4 Postmortem Analyses

Surface chemical analysis of pristine/cycled HC SSEs was performed by XPS in the range of 200–1200 eV, using a VG Scienta SES 2002 spectrometer integrated with a monochromatic X-ray source (Al Kα = 1486.6 eV) and a G Scienta XM780 monochromator. The cycled cells were opened inside the glovebox to recover the electrodes, which were rinsed with DMC and then transferred to the XPS analysis chamber using a sealed box. Besides, SEM and HR-TEM with energy-dispersive X-ray (EDX) mapping analyses of cycled HC SSEs were carried out using the SEM-FEG FEI Quanta 400 and JEOL ARM-200F instruments, respectively.

## 3. Results and Discussion

### 3.1 Physicochemical Characterization

First, the impact of precursor type (three different cellulose-based FPs and a commercial carbon fabric) on the physicochemical properties of the HC SSEs was studied to identify the materials capable of delivering the best electrochemical performance.

The morphological properties of the samples were assessed through SEM analyses. The in-plane SEM images show that all HC SSEs have a similar woven network structure in which 15A, G3, and GA consist of flat fibers stacked in random directions (**Figure 1**a-c), while Kynol consists of multiple bundles of cylindrical fibers stacked in a more orderly manner (**Figure 1**d). As illustrated by the inset histograms (**Figure 1**a-d), the fiber size distribution is different and their average fiber diameter varies widely. Among them, Kynol has the highest average fiber diameter (13 μm), but the narrowest distribution, followed by GA (12.5 μm), 15A (7.4 μm), and G3 (5.8 μm). In addition, the variation in the fiber diameter range of GA is quite significant (2–43 μm) compared to the other HC SSEs (noted in **Table 1**).

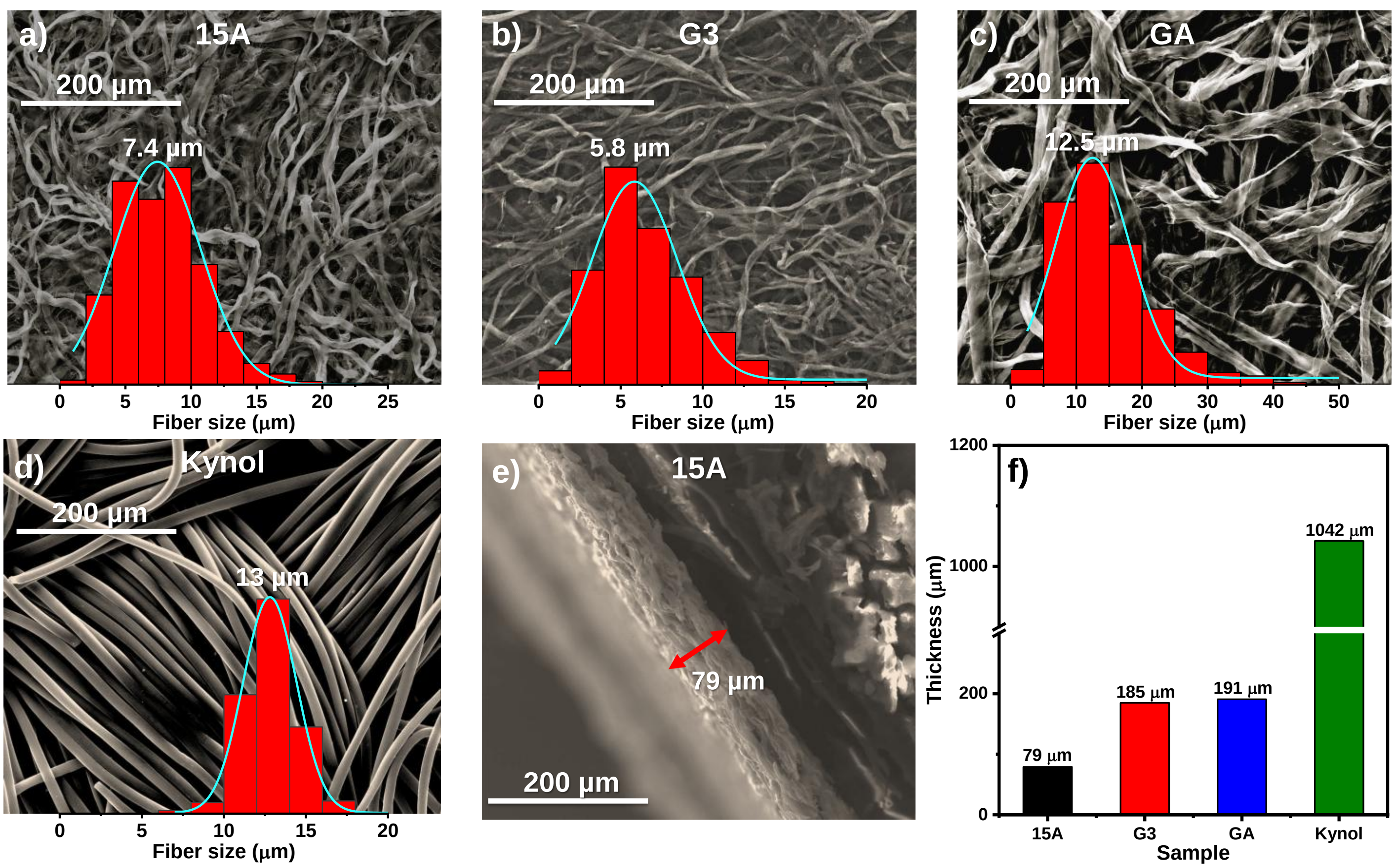

**Figure 1.** In-plane SEM images of HC SSEs: a) 15A, b) G3, c) GA, and d) Kynol. Inset: fiber size (diameter) distribution. e) Cross-section SEM image of 15A and f) thickness of HC SSEs.

**Table 1.** Characteristics of HC SSEs 15A/G3/GA/Kynol, Including the Electrode Thickness, Average Fiber Diameter, Fiber Diameter Range, Mass Loading, Cumulative Pore Volume and Specific Surface Area (SSA $N_2$, SSA $CO_2$)

| Sample | Thickness[a] (μm) | Average fiber diameter[b] (μm) | Fiber diameter range[b] (μm) | Mass loading (mg $cm^{-2}$) | Cumulative pore volume[c] (mL $g^{-1}$) | SSA $N_2$[d] ($m^2$ $g^{-1}$) | SSA $CO_2$[e] ($m^2$ $g^{-1}$) |
|---|---|---|---|---|---|---|---|
| 15A | 79 ± 6.50 | 7.40 | 1–19 | 2.72 ± 0.27 | 2.31 | 37 | 54 |
| G3 | 185 ± 6.41 | 5.80 | 1–17 | 5.46 ± 0.32 | 1.96 | 34 | 41 |
| GA | 191 ± 14.0 | 12.5 | 2–43 | 5.77 ± 0.39 | 3.42 | N/A[f] | 10 |
| Kynol | 1042 ± 29.5 | 13.0 | 7–19 | 29.1 ± 0.41 | 3.19 | N/A[f] | 11 |

[a]Cross-section SEM images. [b]In-plane SEM images. [c]Hg intrusion porosimetry. [d]$N_2$ gas adsorption. [e]$CO_2$ gas adsorption. [f]N/A: The $N_2$ volume is very low to obtain an accurate adsorption isotherm.

By using ImageJ software on the cross-section SEM images of the HC SSEs (**Figure 1**e and **Figure S3, SI**), their thicknesses were determined and presented through the thickness evolution diagram in **Figure 1**f. It can be seen that the thickness order of HC SSEs (**Table 1**)—15A (79 ± 6.50 μm) < G3 (185 ± 6.41 μm) < GA (191 ± 14.0 μm) ≪ Kynol (1042 ± 29.5 μm)—follows the same trend as the precursors (**Table S2, SI**); however, an important decrease was noticed during the pyrolysis process for the three FPs (≈47, 53, and 56%). Meanwhile, the Kynol thickness does not change significantly (≤1%), possibly related to the preheat treatment done by the supplier. The thickness of the electrode is likely to affect the electrochemical performance of the battery, as pointed out in the literature. A large electrode thickness increases the energy density, but there are also limitations such as high mass loading (mg $cm^{-2}$), leading to slow ion and electron diffusion, in addition to high tortuosity and inappropriate porosity.[22,29] The mass loadings of the four HC SSEs are presented in **Table 1**, with Kynol having the highest value (up to 29.1 ± 0.41 mg $cm^{-2}$), followed by GA (5.77 ± 0.39 mg $cm^{-2}$), G3 (5.46 ± 0.32 mg $cm^{-2}$), and 15A (2.72 ± 0.27 mg $cm^{-2}$). Furthermore, due to high internal resistance and poor mechanical integrity of thicker electrodes, their cycling stability is also impaired, as discussed elsewhere.[30] Thus, the different thicknesses and mass loadings observed for the four HC SSEs should allow to establish a better relationship with the electrochemical performance. Regarding the chemical composition of the HC SSEs (**Table S4, SI**), similarities in C content (≈97 wt %) and O content (≈2 wt %) are found among the three FP-based samples. For Kynol, although there is a slight increase in C content and a slight decrease in O content compared to the commercial Kynol carbon fabric precursor (**Tables S3 and S4, SI**), the values are rather different than FP-SSEs. Besides, small traces of Al (0.09–0.3 wt %) were detected in the samples, originating from alumina boats or tubes during the pyrolysis process.

Another particularity of the HC SSEs is the spacing between the stacked fibers (up to hundreds of micrometers) which forms macropores (pseudopores) that can facilitate the transport of $Na^+$ and electrolyte through the interlayer spaces and closed pores for storage, thereby indirectly improving the electrochemical performance of HC SSEs.[31] However, the large pore sizes can also negatively affect the performance, as the contact area between the electrode and the electrolyte increases, leading to unwanted side reactions and higher irreversible capacity during the first cycle.[32] To assess such large pores, mercury (Hg) intrusion porosimetry analysis was used. The obtained Hg intrusion isotherms (lines of solid symbols, **Figure 2**a) are characterized by two main stages: a slope at low pressure region (≤1 MPa), corresponding to the compression of the samples and the penetration of Hg into the spaces between the fibers,[33] and a plateau at the high pressure region (>1 MPa), related to the complete Hg penetration into the macropores. In addition, for the FP-based HC materials, a small shift of 0.1–0.2 MPa can be observed, caused by the transfer of the cells from the low-pressure port to the high-pressure port. Once the device reaches maximum pressure, it begins to decrease and Hg extrusion takes place. A hysteresis between the Hg intrusion and extrusion at high pressure can be observed only for FP-derived HC, especially 15A, which presents the smallest pore size. The main reasons mentioned in the literature are related to the incomplete removal of the Hg due to the complex pore architecture (closed pores or semiclosed pores—like the ink-bottle shape of the pores), contact angle hysteresis, and pore deformation under high pressure.[34]

The cumulative pore volumes of all HC SSEs (**Table 1**), determined through Hg intrusion curves, show the following order: GA (3.42 mL $g^{-1}$), Kynol (3.19 mL $g^{-1}$), 15A (2.31 mL $g^{-1}$), and G3 (1.96 mL $g^{-1}$). Besides, the PSD (**Figure 2**b) reveals that the macropores of the HC SSEs have a very large size range with broad and intense peaks extending from 0.3 to 300 µm. In the region below 10 µm, 15A and G3 have the sharpest peaks, centered at 3.4 and 4.6 µm, respectively. In the next region, from 10 to 100 µm, a sharp and strong peak can be seen at around 12 µm for GA. In addition, a large peak extending from 10 to over 100 µm can be observed for Kynol, with the highest intensity at 107 µm. The presence of these large pores is consistent with the SEM analysis results. In addition to the large pores, smaller ones might be present on the fibers. To check this, $N_2$ and $CO_2$ gas adsorption and pore size distributions (PSD) were determined. The $N_2$ sorption isotherms of 15A and G3 (**Figure 2**c) are typical for low-porosity materials (type II). Their BET specific surface areas (SSAs) are relatively low: 37 $m^2$ $g^{-1}$ for 15A and 34 $m^2$ $g^{-1}$ for G3 (**Table 1**). The PSD (**Figure 2**d) shows a combination of micropores (<2 nm), mesopores (≈2.5–3 nm), and the inter-fiber porosity (≈24 nm) for these two HC SSEs. The micropore volume of 15A (≈0.020 $cm^3$ $g^{-1}$·nm) is higher than that of G3 (≈0.015 $cm^3$ $g^{-1}$·nm), but an opposite trend is seen for the meso- and interfiber pore volumes (**Figure 2**d).

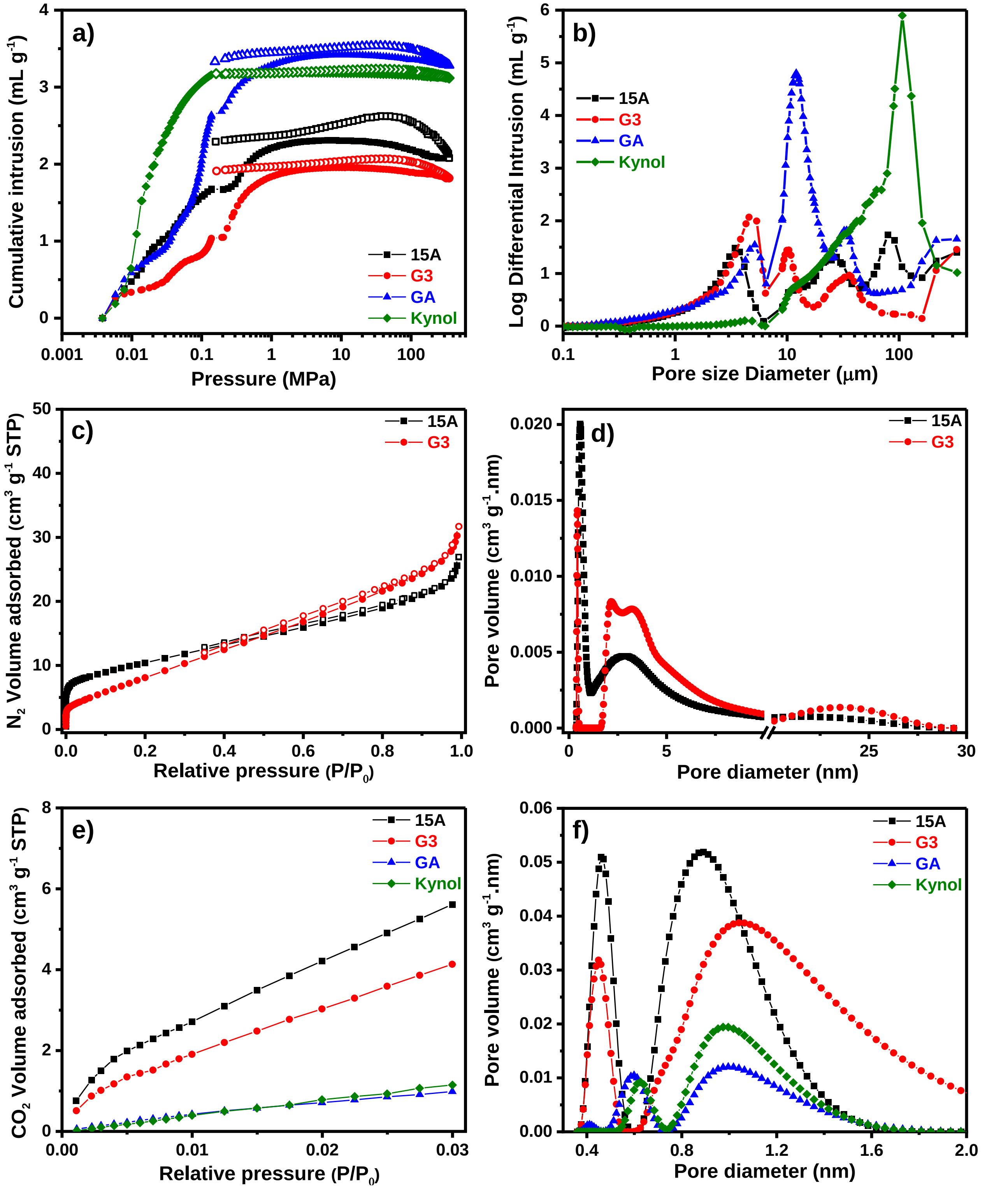

**Figure 2.** Textural characterization of the HC SSEs: logarithmic scale of a) Hg intrusion/extrusion isotherms and b) the corresponding PSD. c) $N_2$ adsorption–desorption isotherms and d) the 2D-NLDFT PSD based on $N_2$ adsorption. e) $CO_2$ adsorption isotherms and f) the 2D-NLDFT PSD based on $CO_2$ adsorption.

It is worth to mention that GA and Kynol samples adsorb very low amounts of gas (within the detection limit) making it impossible to provide an accurate adsorption isotherm, and thus they are not presented. This can be due to the narrow pores that are inaccessible to $N_2$, as demonstrated elsewhere.[35]

In fact, the small molecular size and higher saturation pressure of $CO_2$ compared to $N_2$, provides better accessibility into the narrow pores (<1 nm), called ultramicropores.[35,36] This gas allowed to obtain exploitable isotherms for GA and Kynol, unlike the case of $N_2$. The adsorption isotherms (**Figure 2**e) show that 15A adsorbs the highest volume of $CO_2$, followed by G3, Kynol, and finally GA, with corresponding SSA values of 54, 41, 11, and 10 $m^2$ $g^{-1}$, respectively (**Table 1**). The presence of ultramicropores is also confirmed by PSD which shows bimodal distributions (**Figure 2**f), with sizes centered at ≈0.45/0.9 nm for 15A, ≈0.45/1 nm for G3, ≈0.6/1 nm for both GA and Kynol.

**Table 2.** Characteristics of HC SSEs 15A/G3/GA/Kynol, Including the Interlayer Spacing ($d_{002}$), Crystallite Height ($L_c$) and Crystallite Length ($L_a$), Disorder Degree ($I_{D1}/I_G$), Active Surface Area (ASA), and Oxygen-Based Surface Groups ($CO_x$)

| Sample | $d_{002}$[a] (Å) | $L_c$[a] (Å) | $L_a$[a] (Å) | $I_{D1}/I_G$[b] | ASA[c] ($m^2$ $g^{-1}$) | $CO_x$[c] (mmol $g^{-1}$) |
|---|---|---|---|---|---|---|
| 15A | 3.79 | 11.3 | 36.5 | 2.59 | 1.27 | 0.0517 |
| G3 | 3.81 | 11.2 | 35.4 | 2.59 | 1.94 | 0.0623 |
| GA | 3.76 | 11.6 | 37.0 | 2.47 | 1.46 | 0.0522 |
| Kynol | 3.78 | 8.48 | 39.9 | 2.52 | 0.48 | 0.0230 |

[a]XRD. [b]Raman spectroscopy. [c]TPD-MS.

As a general observation, all HC SSEs developed herein show low SSAs and are quite comparable to other HC powders reported in the literature; for example, a mimosa tannin-based HC (1.5 $m^2$ $g^{-1}$ for $N_2$ / 16.5 $m^2$ $g^{-1}$ for $CO_2$),[37] or a starch potato-based HC (HC-P-St: 2.4 $m^2$ $g^{-1}$ for $N_2$ / 2.3 $m^2$ $g^{-1}$ for $CO_2$).[38] Generally, the higher the pyrolysis temperature, the smaller the specific surface area.[6]

XRD analyses (**Figure 3**a) were performed to provide the structural properties of the HC SSEs. The broad peaks related to (002), (100/101), and (110) graphite planes, characteristic of the disordered hard carbon materials, can be observed in all cases (at 23°, 43°, and 80°, respectively). Among them, the (002) peak of Kynol is much broader and the (100) peak has a higher intensity than the other samples, which may indicate a structure with fewer stacked graphene layers but longer lateral size. Regarding the interlayer distance $d_{002}$ (**Table 2**), G3 has the highest value (3.81 Å) followed by 15A (3.79 Å), Kynol (3.78 Å), and GA (3.76 Å), respectively. Through the Scherrer equation, the pseudographitic domains sizes, $L_c$ and $L_a$, were determined. For the FP-based materials, $L_c$ and $L_a$ were found to be 11.2 Å/35.4 Å for G3, 11.3 Å/36.5 Å for 15A, and 11.6 Å/37.0 Å for GA (**Table 2**). Thus, both $L_c$ and $L_a$ are inversely proportional to $d_{002}$. Based on $L_c$ and $d_{002}$, the number of stacked graphene layers ($N$) was determined to be about 3–4 layers. Meanwhile, Kynol, with the broader (002) peak and the more intense (100/101) peak, has a smaller crystallite height of 8.48 Å (corresponding to ≈3 stacked graphene layers) and a higher crystallite length of 39.9 Å, which is different from the other HC SSEs. Thermal treatment at 1500 °C increased the graphitization degree of the Kynol sample (enhancing the size of pseudographitic domains) as indicated by the higher $L_c$ and $L_a$ values compared to its precursor homologue (**Tables 2** and **S3, SI**). Besides, XRD analysis of representative G3 samples from different batches (**Figure S4, SI**) was used for comparison. The results show overlapping positions of the characteristic peaks of HC, and the $d_{002}$, $L_c$, and $L_a$ values of the samples between different batches are rather similar (**Table S5, SI**). This confirms the reproducibility of the preparation method, as the structural properties of the samples are maintained. Raman analyses of the HC SSEs were performed to provide additional structural information. All Raman spectra (**Figure 3**b) possess two intense and narrow peaks corresponding to the *D* band (≈1350 $cm^{-1}$) and *G* band (≈1600 $cm^{-1}$), typical of HC materials. Furthermore, based on some modification of the Sadezky et al.*'s*[26] procedure, their deconvolution spectra (**Figure S5, SI**) show a main peak *D*1 at 1350 $cm^{-1}$, and also two secondary peaks at around 1200 and 1520 $cm^{-1}$ corresponding to *D*4 and *D*3, respectively. Among them, the *D*4 band originates from $sp^2$–$sp^3$ bonds or C–C and C═C stretching vibrations found in polyenes, and this contribution is the smallest.[39] The *D*1 band represents the breathing mode of $sp^2$ atoms in carbon rings of disordered structure and account for the highest intensity. Moreover, the *D*3 band is related to the amorphous carbon domain (functional groups/organic molecules/fragments).[26] Meanwhile, the *G* band is the result of the stretching mode of C–C bonds in rings and chains of graphite-like structure.[40] No significant differences can be seen among the four HC SSEs, as confirmed by the disorder degree values ($I_{D1}/I_G$): 2.47 for GA, 2.52 for Kynol, and 2.59 for 15A and G3 (**Table 2**). Such high ratios are typical of highly disordered materials, in agreement with the XRD results. With Kynol, there is a decrease in $I_{D1}/I_G$ values compared to its precursor (**Tables 2** and **S3, SI**), which demonstrates that the disordered structure of the sample is more limited after pyrolysis.

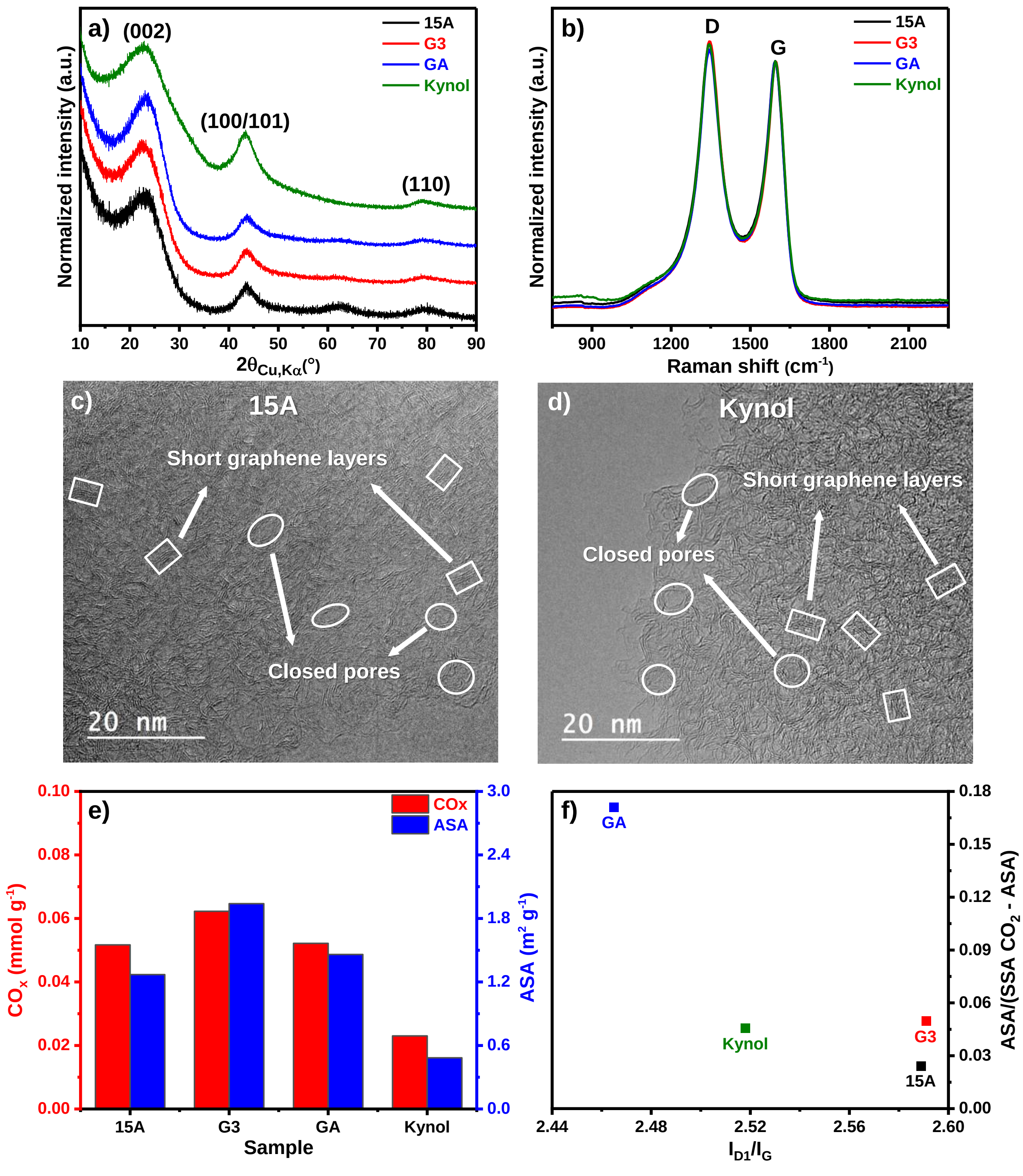

**Figure 3.** Characterization of HC SSEs 15A/G3/GA/Kynol: a) XRD diffractograms; b) Raman spectra normalized based on the G band. HR-TEM images with the presence of closed pores (white oval) and short stacked graphene layers (white rectangle) of c) 15A and d) Kynol. e) $CO_x$ derived from oxygen-based surface groups and active surface area (ASA); f) Graph representing the relation between the ratio ASA/(SSA $CO_2$–ASA) and the disorder degree ($I_{D1}/I_G$).

For more local structural information, HR-TEM analyses were carried out. **Figure 3**c–d and **Figure S6a–b, SI**, show that all materials present a similar and typical structure for HC: highly disordered turbostratic domains consisting of about 3–5 stacked curved and/or straight graphene layers (indicated by white rectangles), in line with the XRD findings. Moreover, closed pores (indicated by white ovals) are significantly present in the structure of all four HC SSEs, which is favorable for $Na^+$ storage.[41]

The nature and quantity of surface functional groups of the HC SSEs were investigated through the $CO_2$ and CO TPD-MS desorption profiles (**Figure S7, SI**). The $CO_2$ desorption gas profile of HC SSEs (**Figure S7a, SI**) shows two broad peaks at about 150 and 310 °C, usually associated with the decomposition of acidic groups, i.e., carboxylic acid (–COOH) and anhydride (–COO–CO), respectively.[38,42,43] At 470 °C, an intense desorption peak is observed only for GA, while at high temperatures, ≈880 °C, broad desorption peaks are present for all materials. The latter are linked to the decomposition of lactone groups.[44] For the CO desorption profile (**Figure S7b, SI**), two broad and low-intensity peaks can be observed at 150 and 350 °C, corresponding to the decomposition of anhydride groups.[38,42] Meanwhile, one can see that CO is the main gas released at temperatures above 600 °C, due to the decomposition of thermally stable functional groups such as ether, carbonyl, and/or quinone.[45] When quantified, these results show the highest amount of $CO_x$ (CO + $CO_2$)—derived oxygen functional groups for G3 (0.0623 mmol $g^{-1}$), followed by GA (0.0522 mmol $g^{-1}$), 15A (0.0517 mmol $g^{-1}$), and Kynol (0.0230 mmol $g^{-1}$), as presented in Figure 3e and Table 2. Interestingly, the $CO_x$ amounts found for these materials are significantly lower (≈10 times) than the values reported in the literature for other HC SSEs prepared at a lower temperature (1300 °C), e.g., 0.66–0.77 mmol $g^{-1}$.[10] Therefore, the higher pyrolysis herein (1500 °C) allows a significant decrease in the O-groups. A small amount of $CO_x$ has been shown to be beneficial for the battery' performance (increased iCE and RC), as mentioned elsewhere.[6] Thus, these HC SSEs are expected to achieve good performance.

The active surface area (ASA) of the materials—comprising edge planes and surface defects, including Stone–Wales (SW) defects, mono-, di-, and/or multiple vacancies, and line defects—was quantified with the help of TPD-MS.[6] The obtained values are presented in **Table 2** and **Figure 3**e. The G3 sample has the highest ASA (1.94 $m^2\ g^{-1}$), followed by GA (1.46 $m^2\ g^{-1}$), 15A (1.27 $m^2\ g^{-1}$), and finally Kynol (0.48 $m^2\ g^{-1}$), with very few defects. The trend is similar to that found for $CO_x$ values. To our knowledge, these ASA values are close to the smallest values of HC powders reported elsewhere, such as HCS1500 (1.2 $m^2\ g^{-1}$), HC-Ch (1.3 $m^2\ g^{-1}$), and HC-L (1.8 $m^2\ g^{-1}$).[38,43] The active surface is an important factor contributing to battery performance, with the presence of defects having both positive and negative effects. Olsson et al.[46] and Youn et al.[47] have shown the positive effect of defect presence in improving sodium insertion and increasing electrochemical performance. However, there are also opposing reviews indicating that a large number of high-energy defects leads to more limited battery performance due to metal trapping.[46] A threshold limit of active sites (4 $m^2\ g^{-1}$) was proposed recently. Below 4 $m^2\ g^{-1}$, the reversible capacity increases with defects due to their favorable interactions with Na-ions, while for higher values, the reversible capacity decays due to the irreversible reactions with the electrolytes.[48] Thus, considering the small ASA values found for the HC SSEs (<2 $m^2\ g^{-1}$), it is likely that their electrochemical performance will be improved.

Lastly, the ratio ASA/SSA $CO_2$–ASA was plotted against the disorder degree ($I_{D1}/I_G$) for the four materials, as illustrated in **Figure 3**f, to provide information on their relationship. As mentioned in the literature, the specific surface area (SSA) is a combination of edge planes/defects (ASA) and basal planes;[13] thus, the ASA/(SSA $CO_2$–ASA) ratio reflects the contribution of surfaces of the active sites (edge planes) to the surface area (basal planes). It can be seen that this ratio tends to increase as the disorder degree decreases. As shown previously, a higher ASA/(SSA $CO_2$–ASA) ratio allows to lower the irreversible capacity and to boost the iCE of HC electrodes made with binders.[13] Regarding the influence of thickness on the intrinsic properties of the material, it can be seen that there is only a correlation between thickness and mass loading. As thickness increases, mass loading increases. Meanwhile, other properties such as morphology, structure, porosity, and surface chemistry are not affected by material thickness, but rather depend on the precursor.[6]

In summary, all FP-SSE materials have quite similar microstructures (interlayer spacing and disorder degree), which slightly differ from Kynol cloth. In addition, low SSA, ASA, and $CO_x$ values are also noted. However, differences can be noticed in the range and average fiber diameter, large porosity (interfiber spacing) between samples, and especially in the significantly higher electrode thickness/mass loading of the commercial carbon fabric compared to the FP-based.

### 3.2 Electrochemical Evaluation

The HC SSEs were electrochemically evaluated, and the results are illustrated in **Figure 4** and noted in **Table 3**. **Figure 4**a shows the first-cycle discharge/charge profiles of the four HC SSEs tested in a half-cell, at a 0.1C rate, using the reference electrolyte (1 M $NaPF_6$ in EC:DMC). Two regions can be observed on the discharge curves, a slope region (of 1.0–0.1 V vs $Na^+$/Na) with a shoulder ranging around 1.0–0.25 V depending on the sample, and a low-voltage plateau region (below 0.1 V vs $Na^+$/Na). Since the $Na^+$ storage mechanism is still under debate and there is no consensus among researchers, a recently proposed mechanism has been considered herein and illustrated in **Figure S8, SI**, for the G3 material.[10,49,50] The $Na^+$ storage during the discharge process occurs as follows: first, the SEI layer formation (shoulder at ≈1.0–0.4 V),[51,52] along with the $Na^+$ adsorption on active sites/open porosity (sloping region from 1.0 down to ≈0.1 V); then, $Na^+$ insertion into graphene layers and/or filling of the closed pores (plateau region, <0.1 V).[53,54]

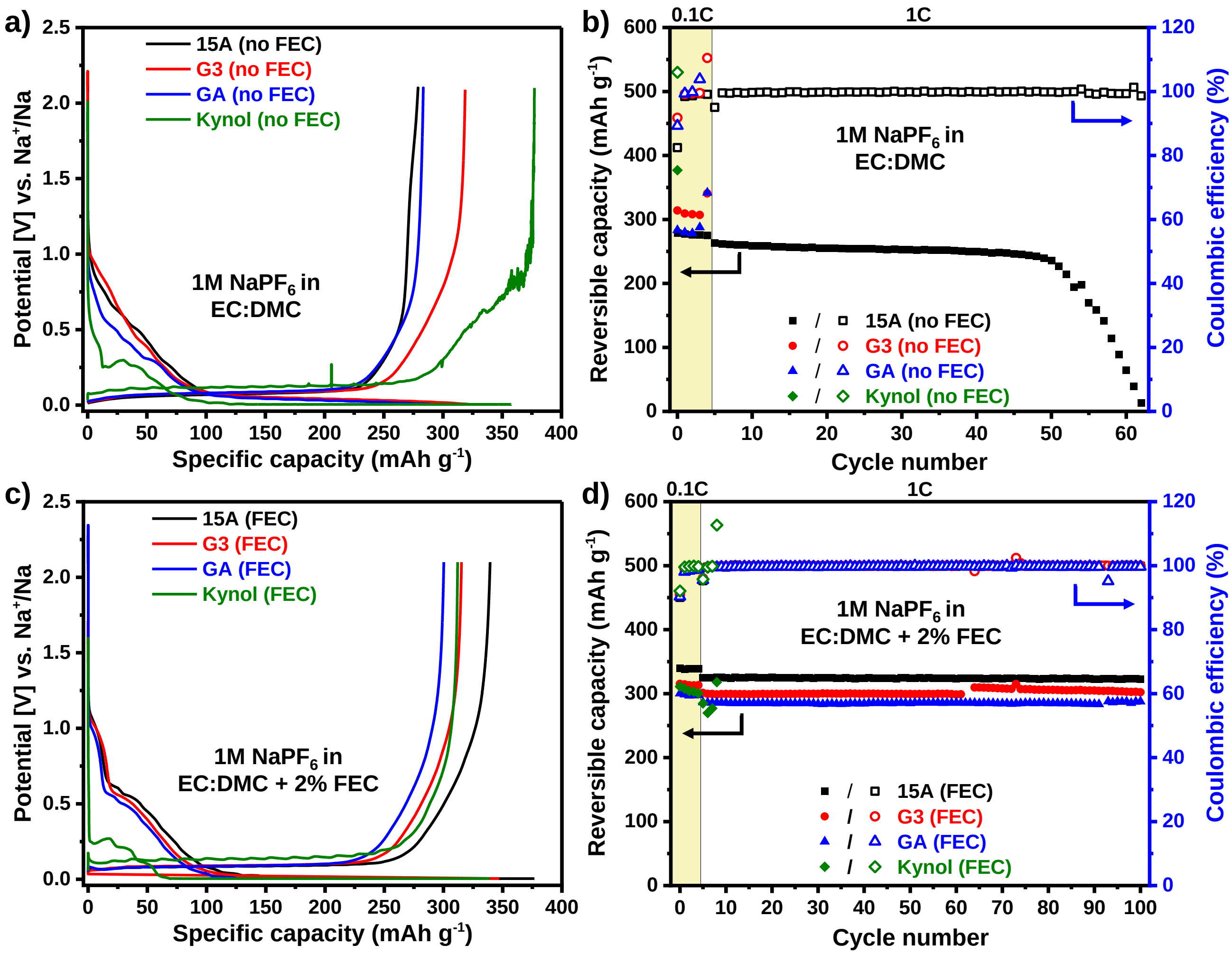

**Figure 4.** Diagrams showing the 1st discharge–charge cycle at 0.1C (left), and the cycling stability (solid symbol) with Coulombic efficiency (hollow symbol) cycled at 0.1C (yellow area) followed by 1C (white area) (right) of HC SSEs 15A/G3/GA/Kynol : a–b) without FEC and c–d) with FEC.

**Table 3.** Electrochemical Performance of HC SSEs 15A/G3/GA/Kynol without and with FEC[a]

| | Sample | iCE (%) | 1st RC (mAh $g^{-1}$) | 1st DC (mAh $g^{-1}$) | Cycling stability | CR (%) |
|---|---|---|---|---|---|---|
| Without FEC | 15A | 84 ± 1.1 | 285 ± 7.1 | 339 ± 9.3 | 50 cycles | X |
| | G3 | 93 ± 0.3 | 319 ± 9.0 | 343 ± 9.1 | 4 cycles | X |
| | GA | 90 ± 0.3 | 283 ± 2.1 | 316 ± 3.5 | 3 cycles | X |
| | Kynol | 105* ± 1.0 | 375* ± 2.1 | 359 ± 1.3 | 1 cycle | X |
| With FEC | 15A | 91 ± 0.4 | 327 ± 7.9 | 361 ± 9.7 | >100 cycles | 99.3 |
| | G3 | 91 ± 0.7 | 313 ± 5.8 | 346 ± 4.4 | >100 cycles | 100 |
| | GA | 91 ± 0.9 | 300 ± 0,5 | 328 ± 3.7 | >100 cycles | 100 |
| | Kynol | 92 ± 0.2 | 322 ± 10.2 | 351 ± 11.9 | 5 cycles | X |

[a]iCE, initial Coulombic efficiency; RC, reversible capacity; DC, discharge capacity; CR, capacity retention; *, performance problem—irregular increase in the reversible capacity.

One can see that the sloping region presents some differences among the four samples. In particular, the SEI formation region is highly dependent on the material type, with Kynol presenting significant electrolyte decomposition reactions compared to the other HC SSEs. The SEI layer is known to be significantly affected by the surface properties (specific surface area, O-functional groups, and defects/active sites).[13] However, the results shows that Kynol presents the smallest $CO_2$ SSA, $CO_x$, and ASA (**Tables 1 and 2**); therefore, such properties are not primarily responsible for its SEI layer properties. It should be noted that this material possesses the highest thickness, fiber diameter, and macropores, which might promote irreversible reactions and inhomogeneous SEI formation. Indeed, Prykhodska et al.[32] observed that a higher macropore fraction contributed to the irreversibility enhancement in polystyrene-templated hard carbon materials. For the other materials, the thinner electrode (≈5 times) led to the formation of a distinct and more homogeneous SEI passivation layer, which affected the iCE and RC differently. Among the HC SSEs, G3 delivers the best performance (**Table 3**) with an impressive iCE of 93%, one of the highest reported for a HC SSEs (**Table S1, SI**), and a RC up to 319 mAh $g^{-1}$, followed by GA (iCE = 90%, RC = 283 mAh $g^{-1}$) and 15A (iCE = 84%, RC = 285 mAh $g^{-1}$). On the other hand, Kynol shows doubtfully high performance (375 mAh $g^{-1}$ for RC and iCE >100%) likely related to the unusual electrochemical signal observed in the charge sloping region (important fluctuations).

The higher performance of the G3 electrode in the first cycle can be related to its properties compared to the other two HC SSEs. It presents the highest $d_{002}$ (3.81 Å) and $I_{D1}/I_G$ (2.59) and might promote the $Na^+$ insertion. These values as well as $CO_x$-based groups and ASA are closed to the other HC SSEs (15A and GA) (**Table 2**). However, the G3 is distinguished from the other materials by its smallest macropore volume (G3 < 15A < GA, **Table 1**). The G3 electrode thickness/fiber diameter (15A < G3 < GA) and the SSA $CO_2$ (GA < G3 < 15A, **Table 1**) fall between those of the other two samples. Based on our previous studies, the relationship between HC properties and iCE has been established, with threshold limits of SSA $CO_2$ (<50 $m^2$ $g^{-1}$), $CO_x$ (<0.25 mmol $g^{-1}$), and ASA (<5 $m^2$ $g^{-1}$) to ensure improved material performance.[13,36,48] Considering these small values of HC SSEs (**Tables 1** and **2**), the materials have the potential for high iCE (≥85%), as demonstrated by the electrochemical results (**Table 3**). Regarding the cyclability (**Figure 4**b), poor stability can be seen for G3, GA, and especially Kynol. Meanwhile, 15A shows very good stability (at both 0.1C and 1C rates) up to 50 cycles, followed by a sudden drop in capacity. As the physicochemical properties of the materials are rather similar, the reason for this performance instability could be highly related to the electrode thickness and morphology (interwoven-fiber structure). The high thickness of the electrode along with the large fiber diameter and interfiber spacing leads to inhomogeneous reactions through the electrode depth. The SEI byproducts accumulate more on the surface and contribute to kinetic limitations and performance decay.[55] It can be seen that the electrode thickness and mass loading are the most impactful parameters, while macroporosity has less impact, in the case of using reference electrolyte (**Figure S9, SI**). For example, G3, which presents the smallest macropore volume and intermediate thickness, faces severe fading (after only 4 cycles) compared to 15A (after 50 cycles), which has a higher macropore volume and smaller thickness (**Table 1**).

Therefore, the thinnest electrode with smaller diameter fibers, 15A, has the best stability, which then deteriorates with increasing electrode thickness and fiber diameter: G3, GA, and Kynol. It appears that the large electrode thickness (185–1042 μm) tends to form uneven and unstable SEI layers due to the promotion of inhomogeneous reactions through the electrode depth which have a negative impact on cycling stability. On the contrary, appropriate thickness (79 μm) seems to lead to a more homogeneous and/or thinner SEI layer which is more stable and can last longer.

The electrochemical tests in the presence of FEC show slight changes in the shape of the discharge curves of the materials (**Figure 4**c), making them slightly rougher than those without FEC (**Figure 4**a). With FP-derived HC SSEs, a shoulder (related to electrolyte decomposition and SEI layer formation) can be clearly observed in the voltage range of 0.7–0.55 V, while for Kynol it is located at ≈0.25 V. This indicates that more electrolyte decomposition reactions take place for HC SSEs when using the FEC additive. The electrochemical performance delivered was also impacted by the presence of FEC (as shown in **Table 3**). G3 shows a slight change in performance: iCE decreases from 93% to 91% and RC drops by ≈6 mAh $g^{-1}$, while 15A and GA HC SSEs show some improvements. For 15A, the iCE enhances significantly, from 84% to 90%, and RC increases by almost 42 mAh $g^{-1}$, while GA presents a gain in RC of 17 mAh $g^{-1}$ (iCE remains constant). More importantly, Kynol can achieve excellent performance now: an iCE of 92% and an RC of 322 mAh $g^{-1}$, without seeing the presence of any fluctuations, thus confirming the positive role of FEC. In addition, when changing the C-rate from 0.1C to 1C, only a small drop in capacity is noticed for all samples (i.e., 11–15 mAh $g^{-1}$).

Besides, the use of the FEC additive significantly improves the cycling stability of the electrodes (**Figure 4**d). Indeed, the FP-derived HC SSEs show very good stability for at least 100 cycles at 1C rate, with RC being maintained at 320 mAh $g^{-1}$ for 15A and ≈290 mAh $g^{-1}$ for G3 and GA, having a capacity retention between 99% and 100% (**Table 3**). However, the cycling performance of Kynol is still limited as it can only perform stably for the first five cycles at 0.1C. Once switching to 1C, problems occur during the charging step once again. Although it is clear that FEC brings improvements to the electrode, its high thickness seems to still limit the formation of a high-quality SEI, as it can only be stable for five cycles before fluctuation takes place. Nevertheless, the ability of the FEC additive to improve the electrochemical stability of the electrodes was confirmed.

The diagram analyzing the RC decay rate per cycle of the three FP-based HC SSEs at 1C, from the 7th to 100th cycle, is presented in **Figure S10a, SI**. The results show that the capacity differences between cycles are negligible, fluctuating within ±1%. This indicates the stability of the samples over 100 cycles when using the FEC-based $NaPF_6$ electrolyte. The symmetrical current density regime tests were performed to investigate the cycling stability of the materials at high currents **Figure S10b SI**. The results indicated that all three materials performed very well and remained stable up to 5C, with the obtained RC of 285–300 mAh $g^{-1}$ and a Coulombic efficiency (CE) close to 100%. At the moment of changing the C-rate, CE shows a slight decrease, as the kinetic reaction of the $Na^+$ insertion/discharge process in the cell is affected by variations in current density. After a few cycles at 10C, significant capacity fluctuations are observed with G3 and GA, with random CE values exceeding 100%. Performance stability is only restored to its original values, with reversible capacity (≈300 mAh $g^{-1}$) and CE (100%), after several cycles when returning to 1C. Meanwhile, HC SSE 15A shows perfect stability throughout the test. This may be related to the difference in thickness of the HC SSEs, as their morphological, structural, and surface chemical properties are quite similar. When the thickness is high (G3/GA ≈ 200 μm), it leads to slower diffusion of $Na^+$ through the spacing between the stacked fibers of the material, or increased polarization, thereby causing cycling instability during rapid discharge/charge.[56] Kynol with very high thickness, showing fading after only a few cycles at 1C (**Figure 4**), could not be tested for higher C-rates.

Compared with the performance of HC SSEs reported in the literature (**Table S1, SI**), one can observe that the materials developed here are among those that provide the best performance, with very high iCE (≥91%), RC (≥300 mAh $g^{-1}$), and CR (≥95%), while applying a higher current density, 37.2 mA $g^{-1}$ vs 10–20 mA $g^{-1}$, typically. Moreover, most of the materials here possess significantly higher mass loadings than other HC SSEs (5–6 mg $cm^{-2}$ vs 1–3 mg $cm^{-2}$), which is an important issue for real applications.

### 3.3 Study of the Solid Electrolyte Interphase (SEI)

To better understand the effect of electrolyte type (without/with FEC additive) on the performance of HC SSEs, XPS analysis of G3 (pristine and after the fifth discharge/charge cycle, without/with FEC) was carried out, through which the surface composition information on the SEI layers was determined. Besides, it was compared with the XPS results of 15A (pristine and after the fifth discharge/charge cycle, with FEC) to gain insights into the impact of the precursor type on the resulting SEI layers.

XPS analysis of the two pristine HC SSEs 15A and G3, providing information on their surface chemical environments before cycling, is presented in **Figure S11, SI**. The results show a rich carbon content in the composition of both materials, consisting of 96.60 at % for 15A and 96.57 at % for G3 (**Figure S11a, SI**), which is beneficial in terms of improving the electronic conductivity of the materials. Additionally, a small percentage of O can be observed, 2.50 at % and 2.55 at %, respectively, thereby limiting the impact of this element in promoting unwanted reactions with the electrolyte.[6] Besides, some traces of impurities (<0.6 at %), possibly stemming from analytical tape or the pyrolysis boat, are noticed. By observing their deconvoluted C 1s XPS spectra (**Figure S11b-c, SI**), it is possible to recognize the dominant distribution of $sp^2$ carbon species (≈87 at %) with intense peaks at around 284.4 eV. In addition, some weak signals can be seen in the region 290–284 eV coming from oxygen functional groups such as C═O (carbonyl), C–OR (ethers), and COOR (carboxylic acid and anhydride groups), with low content (<1.2 at %), in agreement with the small amount of $CO_x$ derived O-groups from the TPD-MS.

The deconvoluted XPS C 1s, Na 1s spectra, and pie chart depicting the surface composition of HC SSEs G3 (without/with FEC) and 15A (with FEC) after the first five cycles at 0.1C are shown in **Figure 5**a–c, respectively. In addition, the XPS spectra of the other two elements (O 1s and F 1s) are provided in the **Supporting Information Figure S12, SI**.

After cycling, the absence of C $sp^2$ peak in the C 1s spectra (**Figure 5**, left), the presence of new species in the other spectra of Na 1s, O 1s, and F 1s (**Figure 5**, middle and **Figure S12, SI**), and the contribution of Na, P, and F elements in the surface composition (**Figure 5**, right) indicate that the formed SEI layer covered the electrode surface.

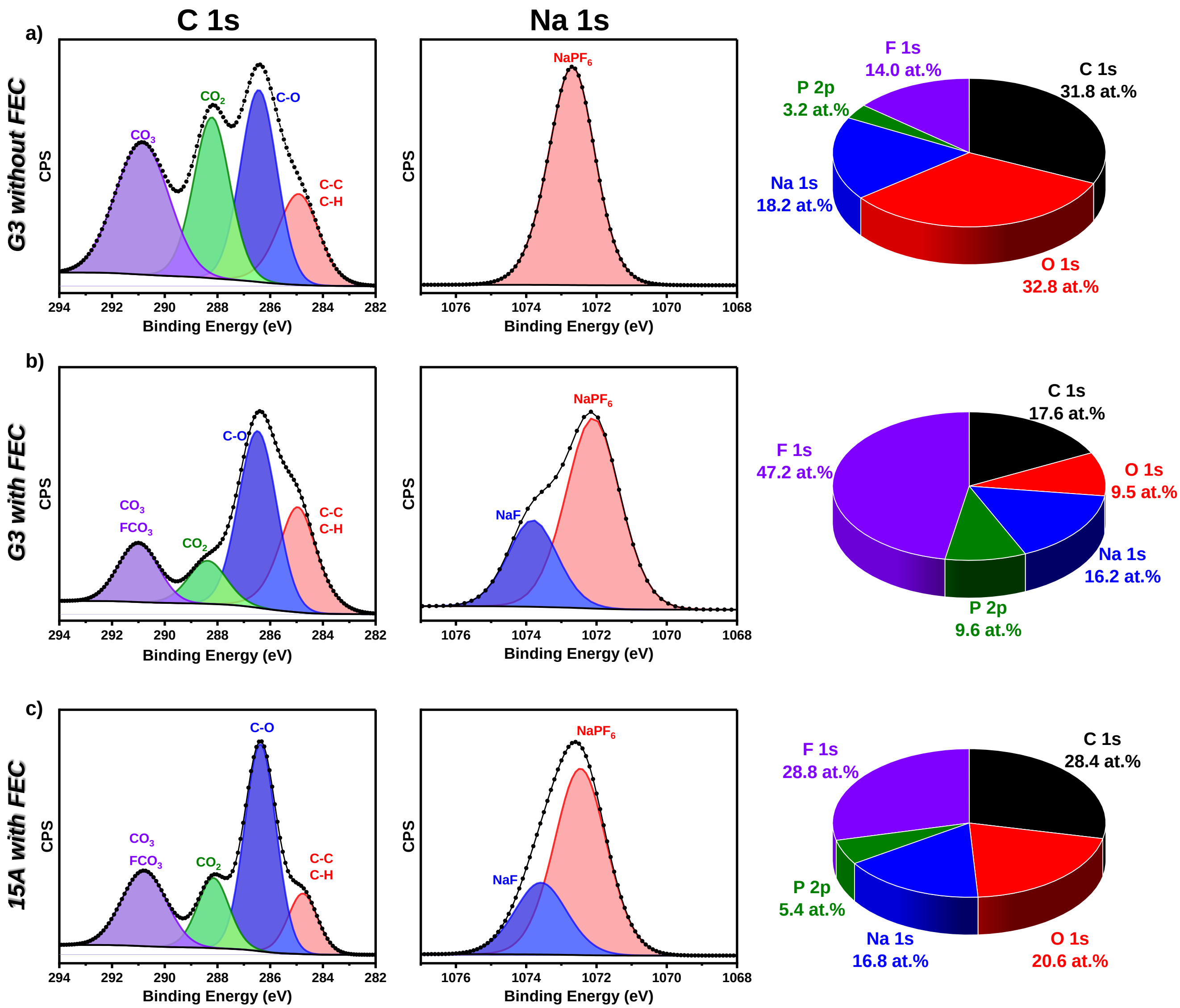

**Figure 5.** Deconvolution of XPS peaks—C 1s (left) and Na 1s (middle)—and surface composition (right) of SSEs cycled at the fifth discharge/charge cycles: a) G3 without FEC, b) G3 with FEC, and c) 15A with FEC.

Going into more detail, the presence of four distributions in the C 1s spectra (**Figure 5**, left) can be seen. Among them, the most intense peak, in the binding energy (BE) range 286.50–286.32 eV, is characteristic of C–O bonds in poly(ethylene oxide)—PEO $(–CH_2–CH_2–O)_n$ and sodium alkyl alkoxides (RONa), two of the decomposition products of EC and DMC solvents.[57] Next to it, there is a peak in the BE in the range of 285.0–284.7 eV, corresponding to the C–C and/or C–H bonds, which are associated with aliphatic chains.[58] At higher BE, there are also two other notable peaks located at ≈288 eV and ≈291 eV, corresponding to the $CO_2$ species (O–C═O bond) and $CO_3$ species (with $FCO_3$ species—in case of using FEC), respectively. $CO_2$ originates from sodium alkyl carboxylate ($RCO_2Na$), while $CO_3$ is associated with sodium carbonate ($Na_2CO_3$) and/or sodium alkyl carbonates ($ROCO_2Na$).[57] Indeed, these CO/$CO_2$/$CO_3$ species signals, characteristic of C–O and C═O bonds, can also be observed with a high-intensity peak at around 532 eV in the O 1s spectra (**Figure S12,** left, **SI**). Besides, sodium ethylene dicarbonate ($NaO_2CO–C_2H_4–OCO_2Na$), formed by the decomposition of the EC solvent, has also been mentioned as the main constituent of the SEI layer.[59] Looking at the distribution, the $CO_2$ and $CO_3$ peak intensities decrease significantly when FEC is used, which corresponds to the diminution of organic compounds making up the SEI layer. Additionally, 15A presents more SEI compounds than G3, especially organic species (enhanced peak intensity of CO, $CO_2$ and $CO_3$), likely due to the higher SSA $CO_2$ and macropore volume (**Table 1**) leading to more electrolyte decomposition reactions.

For all three cycled HC SSEs, the Na 1s spectra possess a high-intensity peak at a BE of 1072.71–1072.16 eV, representing the Na–F bond of the $NaPF_6$ salt (**Figure 5**, middle).[24] In addition, with the use of FEC, the NaF signature can be observed at 1073.76–1073.60 eV (**Figure 5**b–c, middle).[60] Furthermore, the F 1s spectra confirm the presence of NaF, which is accompanied by the appearance of the $NaPF_x$ peak at around 685 eV, next to the intense peak at 688 eV—characteristic of the $NaPF_6$ salt and the C–F bond, (**Figure S12**, right, **SI**).[60,61]

Looking at the surface composition of G3 using the electrolyte without FEC (**Figure 5**a, right), it can be seen that the C and O contents are predominant (>60 at % in total) compared to the other elements—Na, P, and F. This shows that the main components of the SEI layer are organic (PEO, RONa, $RCO_2Na$, $ROCO_2Na$) with a minor part of inorganic ($Na_2CO_3$ and $NaPF_x$) species.

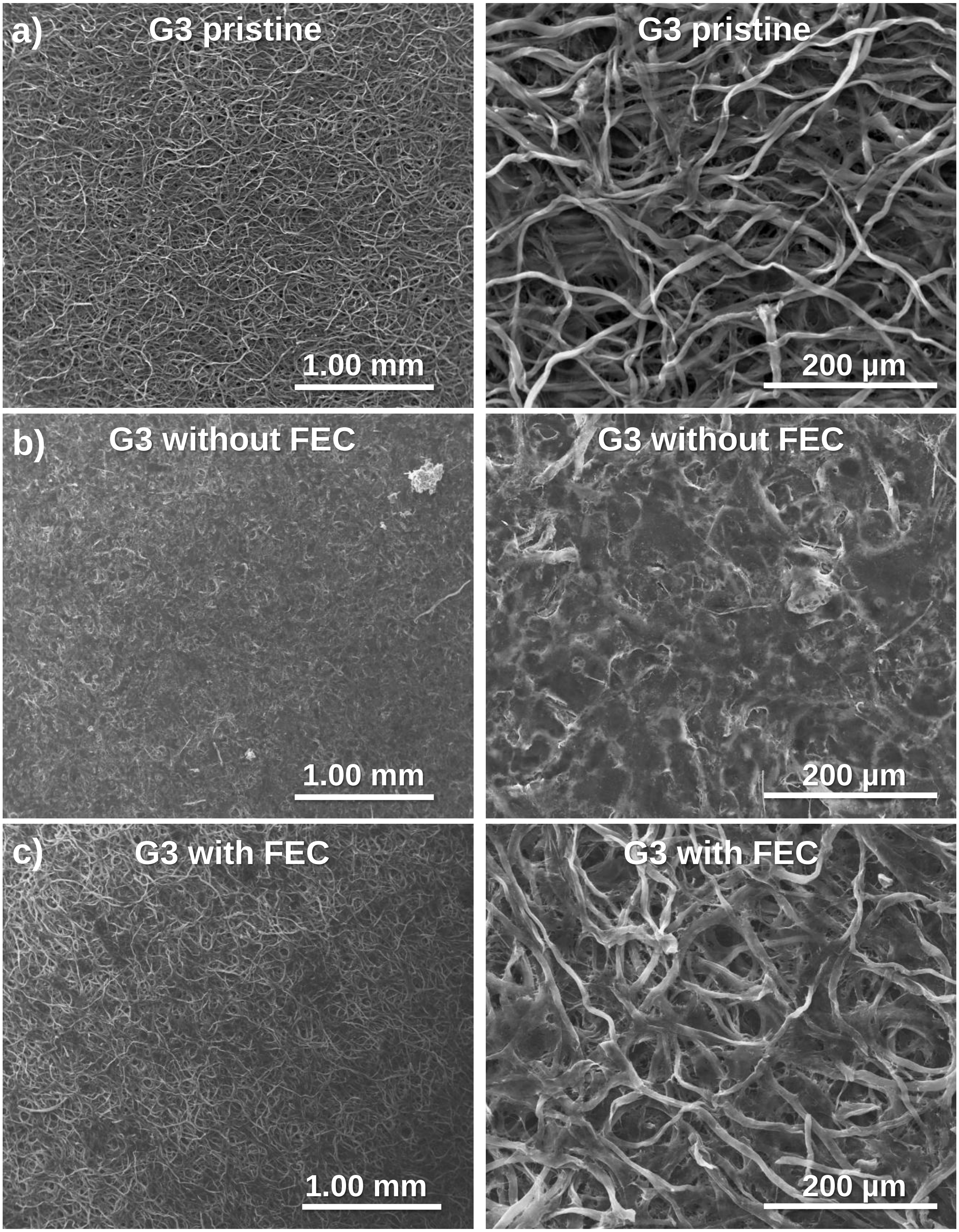

**Figure 6.** SEM images of HC SSEs G3: a) pristine, and after cycling: b) without FEC and c) with FEC.

With the use of FEC additive, the surface chemical composition of G3 shows a clear transformation with a significant decrease in the C (from 31.8 to 17.6 at. %) and O (from 32.8 to 9.5 at. %) content (**Figure 5**b, right), in line with the intensity decrease of the $CO_2/CO_3/FCO_3$ peak in the C 1s spectrum (**Figure 5**b, left). In addition, an important content of F (47.2 at. %) and a slight increase in P content (from 3.2 to 9.6 at. %) can be noted (**Figure 5**b, right), which can be linked to the formation of NaF, as well as the enhanced presence of $NaPF_x$, respectively.

Besides, it can be found that the SEI layer of 15A is built up with a more balanced composition between the organic/inorganic species ($–CO/–CO_2/–CO_3/–FCO_3$ or $Na_2CO_3$) and fluorinated inorganic species of ($NaF/NaPF_x$) (Figure 5c, right), compared to G3, which contains more inorganic species (NaF, 47.2 at. %). This can be due to the higher SSA and macropore volume of 15A sample compared to G3 (**Table 1**).

In short, the XPS results show that in the absence of FEC in $NaPF_6$ electrolyte, an organic-rich SEI layer (Na alkyl-alkoxide/carboxylate/carbonate) with a small amount of inorganic species ($NaPF_x$, $Na_2CO_3$) is formed on G3 SSE. The organic-rich SEI layer is known to be thick, uneven, unstable, and prone to dissolution in organic electrolytes. This leads to continuous regeneration and electrolyte consumption, resulting in a reduction of performance.[6] In the presence of FEC, the SEI layer composition changes, becoming richer in NaF inorganic species, especially in greater quantities in G3 compared to those in 15A. The NaF was reported as very efficient to improve the SEI layer properties. It leads to a more robust and stable SEI that provides good ion conduction and limits the further decomposition of the electrolyte.[62] Furthermore, FEC can form a protective film on the counter/reference electrode surface, preventing unwanted decomposition reactions of unstable carbonate-based electrolytes on highly reactive Na metal, as indicated in the literature.[63,64]

To get more information on the SEI layer, SEM images were performed on G3 electrodes after cycling in both electrolytes (without/with FEC) and compared to the pristine electrode (**Figure 6**). Significant morphological changes can be noticed as the individual fibers of the pristine electrode (**Figure 6**a) are now covered/glued by a dense film for the cycled electrode without FEC additive (**Figure 6**b). This film can correspond to the SEI layer, which, due to its thick aspect, can significantly limit $Na^+$ kinetics (insertion/extraction) and degrade the performance during cycling, consistent with the electrochemical results found. Very interestingly, when FEC is present, the SEI formation appears to be significantly limited, since the fibers can be clearly seen now, even if they are connected to each other through a film on certain areas. The SEI layer seems to rather surround the fibers, as their thickness slightly increased (**Figure 6**a,c), which proved to be more favorable for the electrochemical behavior.

Considering these observations, the other electrodes were evaluated as well after cycling while using electrolytes with FEC (**Figure S13, SI**). Similar findings were noticed for 15A, GA, and Kynol electrodes. More precisely, the individual and smooth fibers are now covered by a thin SEI film which extends between the fibers for some samples (15A and GA) or presents small particles on their surface, as it is the case for Kynol (**Figure S13d, SI**). These particles maybe at the origin of limited cyclability of Kynol electrodes.

To obtain further insights, the local morphology and chemical composition of the cycled G3, using FEC-based $NaPF_6$ electrolyte, was evaluated by HR-TEM (**Figure 7**a) and EDX analysis (**Figure 7**b–e and **Figure S14a–d, SI**). Compared with the TEM image of pristine G3 (**Figure S6a, SI**), there is a significant structural difference, as disordered turbostratic stacked graphene layers are not visible for the cycled G3. The zoomed-in image presented in the inset reveals a continuous and porous layer covering the particle surface.

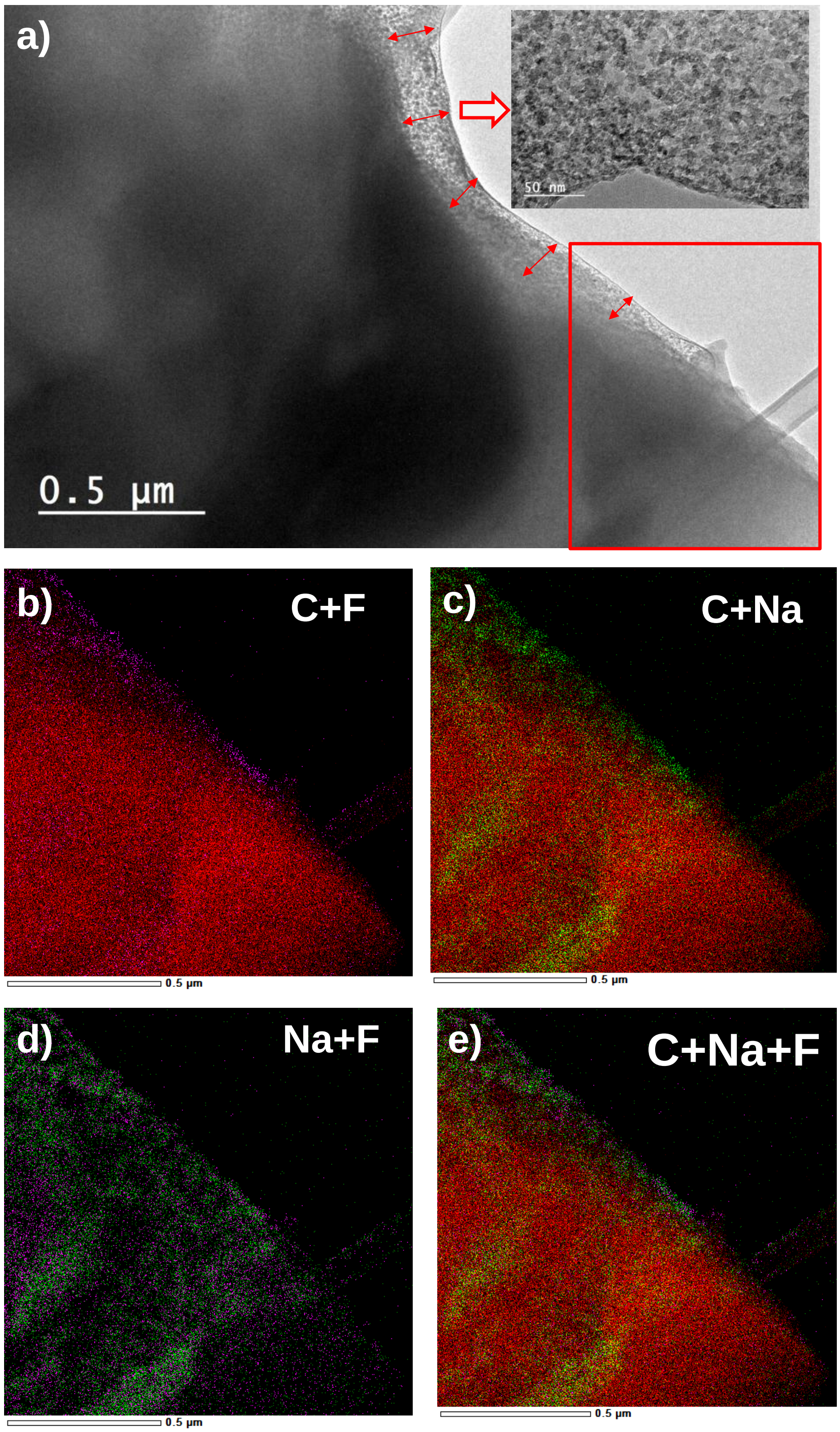

**Figure 7.** (a) HR-TEM image, and EDX mapping of G3 cycled for 5 cycles in FEC-based $NaPF_6$ electrolyte showing: b) C and F overlapping, c) C and Na overlapping, d) Na and F overlapping, and e) C, Na, and F overlapping, where C is in red, F is in purple, and Na is in green.

This corresponds to the SEI layer, as confirmed by EDX showing a fairly homogeneous presence of C, Na, F, and O elements in the G3 structure (**Figure 7**b–e and **Figure S14a–d, SI**). Furthermore, the distribution of Na and F elements is found at the edge of the HC SSE (**Figure 7**b–c) and on the entire material surface (Figure 7d). Furthermore, on the overlapped C + Na + F, one may see that the C is found predominantly on the material surface and Na + F in the deeper layer. This suggests that the SEI layer is composed of an inorganic-rich (NaF) core and an organic-rich surface layer.

To clarify the impact of SEI layer on the resistance, electrochemical impedance spectroscopy (EIS) was conducted on three samples similar to the XPS analysis (G3 without FEC, G3 with FEC, and 15A with FEC). Three conditions analysis were investigated as follows: before cycling (freshly assembled cells), after the first cycle (newly formed SEI layer), and after five cycles (stable SEI layer) at 0.1C. Based on a report by Lu et al.,[65] the EIS data of the three samples before cycling were fitted and matched to the same equivalent circuit (**Figure S15, SI**). This circuit was set up consisting of R1 (intersection on the real axis—bulk resistance) connected in series with the parallel circuit Q2//R2 (semicircle—SEI layer), Q3//R3 (charge transfer across the electrode surface) then W4 (Warburg impedance), and finally Mg4 (abnormal diffusion when the Warburg impedance is not a straight 45° line).

Before cycling, it can be seen that the Nyquist plots of all three samples (black traces) start very close to 0 at around 2–3 Ω, which indicates that the bulk resistance (R1) of the cells (relative to the electrodes/electrolyte/separators) is very low (**Figure 8**a–c). In addition, they consist of two main parts: a well-defined semicircle at high frequencies and a straight line at low frequencies. These semicircles are usually associated with the formation of the SEI layer and the charge-transfer resistance.[66]

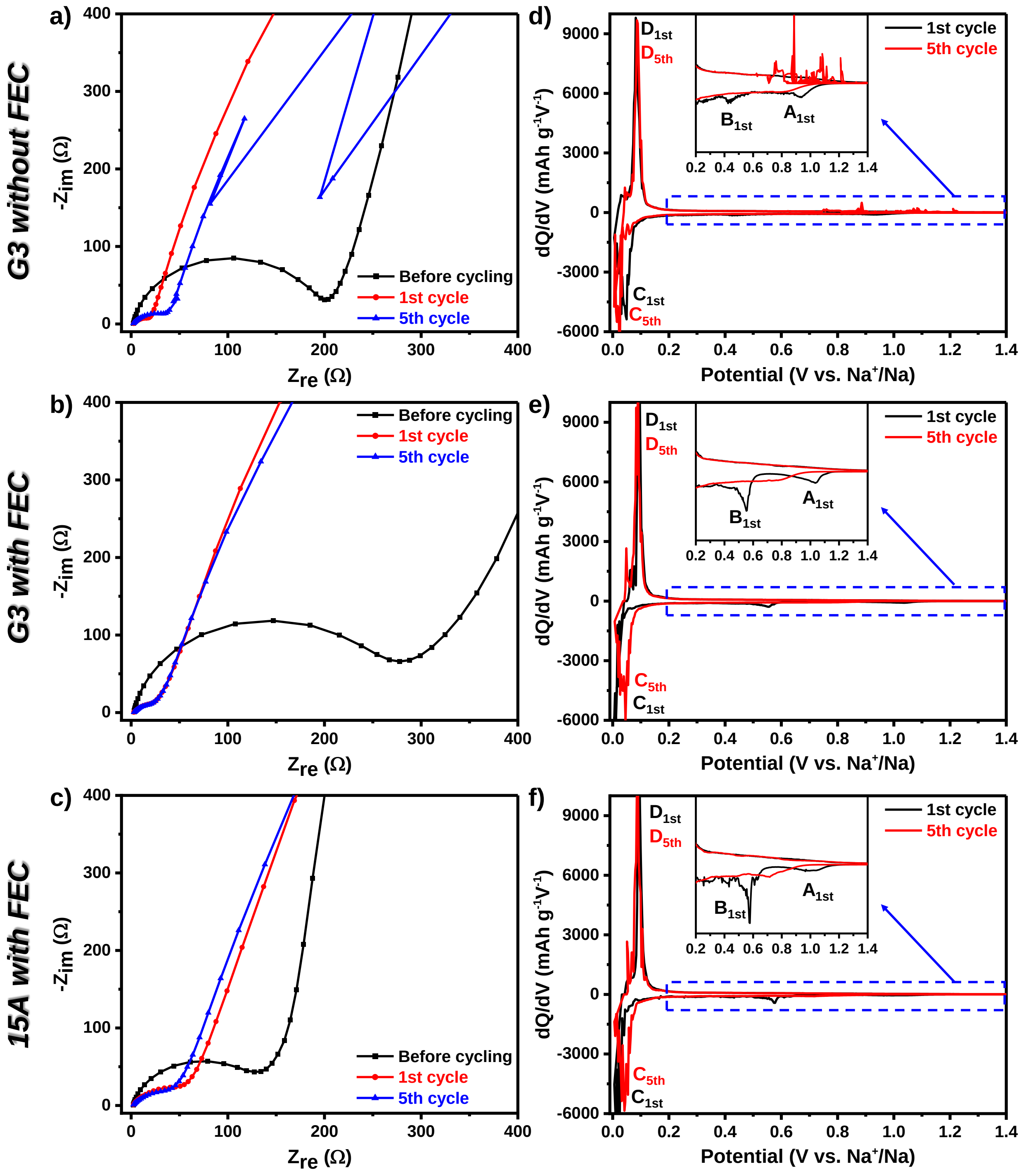

**Figure 8.** Electrochemical impedance spectra of HC SSEs: a) G3 without FEC, b) G3 with FEC, and c) 15A with FEC, measured before electrochemical testing (black trace), after testing the material at 0.1C for 1 cycle (red trace), and for 5 cycles (blue trace). Differential capacity (dQ/dV) vs voltage (V) plots with the zoomed-in images of the voltage region 0.2–1.4 V (inset) of HC SSEs: d) G3 without FEC, e) G3 with FEC, and f) 15A with FEC, cycled at 0.1C at the first cycle (black trace) and fifth cycle (red trace).

This suggests that as soon as the cell components (HC/electrolyte/Na) come into contact, the electrolyte decomposition reactions are immediately promoted to form the passivation layer, consistent with recent research.[24] More specifically, this layer is first built up by the deposition of some inorganic species ($Na_2CO_3$, $NaPF_x$, including NaF when using FEC) on the carbon surface through the electrolyte decomposition reaction. The straight lines in the low-frequency region are usually linked to the Warburg impedance, which characterizes the diffusion of $Na^+$ into the electrode.[66] Regarding the resistance, it can be noted that the G3 without FEC (R2 ≈ 200 Ω) has a smaller value as the G3 with FEC (R2 ≈ 278 Ω). The higher resistance of G3 in the presence of FEC may be related to the higher NaF content, the latter being known to have a higher resistance than $Na_2CO_3$.[67] In addition, 15A cycled in FEC-based electrolyte shows a significantly smaller resistance (R2 ≈ 127 Ω) than the other two samples. This is explained by its significantly smaller thickness (79 µm) compared to G3 (185 µm),[68] as well as its more balanced SEI distribution between NaF/$Na_2CO_3$, rather than NaF-rich in G3 with FEC, with the first factor showing a more important impact on the resistance before the cycling.

After the first cycle, the impedance results of all three cells (red traces, **Figure 8**a–c) show a contraction of the semicircles, as well as a significant decrease in resistance, with G3 without FEC giving the smallest value (≈ 17.5 Ω), followed by G3 FEC (≈ 20 Ω) and 15A FEC (≈ 50 Ω). This shows that the passivation layers have formed quite completely and homogeneously; the HC surface is covered with an inorganic layer followed by an organic layer in contact with the electrolyte surface. The $Na_2CO_3$-rich SEI layer of G3 (without FEC) provides the best conductivity and the highest iCE (93%) compared to of the other two samples, which have NaF-rich SEI. In addition, due to the larger SSA $CO_2$ and macropores of 15A, the electrochemical cycling promotes a more intense electrolyte decomposition, forming a thicker SEI layer resulting in a higher internal resistance than G3 when using FEC.

However, after the first five cycles, the resistance of G3 without FEC suddenly increases as well, as its Warburg line is somewhat distorted (blue trace, **Figure 8**a). This result is associated with the fluctuation of the reversible capacity of this sample. This negative change in resistance indicates the instability of the SEI layer when it is too rich in $CO/CO_2/CO_3$ species (high C and O atom %, observed in XPS), hindering the diffusion of $Na^+$. Meanwhile, with the two samples using FEC, the semicircles and resistances decrease very slightly with G3 and moderately with 15A (blue traces, **Figure 8**b–c), suggesting that the SEI layer is improved, more stable, and homogeneous. This reinforces the positive effect of FEC in stabilizing the SEI layer, thereby enhances the battery cycling stability.

To further elucidate the relation between the electrochemical behaviors and the SEI layer, the differential capacity (dQ/dV) vs voltage (V) plots of three cells: G3 without FEC/G3 with FEC/15A with FEC, cycled at 0.1C during the first cycle and the fifth cycle, are used for investigation and are presented in **Figure 8**d–f. All cells have two distinct voltage regions for sodiation/desodiation process: the sloping region (1.2–0.15 V) and the plateau region (0.15–0.005 V).

In the sloping region of the first cycle (black traces), two weak and broad sodiation peaks ($A_{first}$ and $B_{firts}$) can be observed at 0.94 V–0.43 V for G3 without FEC (**Figure 8**d, inset). These signals are usually associated with the $Na^+$ adsorption on the active sites and the decomposition of the electrolyte to form the SEI layer.[69] In the presence of FEC, these processes are promoted on G3 and 15A as their peak shift to the higher potential region at 1.03 V–0.55 V and their intensity increases especially for the peak at 0.55 V. This can be linked to the enhanced decomposition of the FEC additive and $NaPF_6$ electrolyte to form the inorganic-rich SEI layers, as observed in both the first discharge/charge curve (via the well-defined shoulder, **Figure 4**c) as well as the EIS results (**Figure 8**b–c).

In contrast, the desodiation process of the three samples is quite similar, and no peak is recorded. At the fifth cycle, a clear decrease of the sodiation peaks can be seen (red traces, **Figure 8**d–f, inset), which indicates that the formed SEI layer has stabilized. However, fluctuation in the signal of G3 without FEC occurred when the potential of the desodiation increased from 0.7 V (red trace, **Figure 8**d, inset). This problem is reflected in the electrochemical and impedance results. Due to the impact of material properties such as high fiber diameter, interfiber spacing, and especially large thickness, which promote heterogeneous reactions through the electrode depth, a thick, inhomogeneous, unstable organic-rich SEI layer is formed (confirmed by XPS), thereby leading to kinetic limitations (confirmed by EIS) and performance degradation through cycles.[55]

In the plateau region, sharp and intense peaks of sodiation—$C_{first}$ (0.07–0.005 V)/desodiation—$D_{first}$ (0.005–0.12 V) can be observed for all cells. These peaks are usually linked to the insertion/extraction of $Na^+$ into/out of the disordered graphene layers or/and closed pores, respectively. By comparing the first and fifth cycles, it can be noticed that the hysteresis of the sodiation/desodiation peaks (distance between $C_{fifth}$ and $D_{fifth}$ peaks, **Figure 8**d) of the G3 without FEC increases, while it decreases for G3 and 15A using FEC (**Figure 8**e–f). The hysteresis rise reflects an increase in cell resistance,[70] thus slowing down the kinetics of Na-ion insertion/extraction in the structure of G3 without FEC, which has been demonstrated in its impedance analysis. As a result, the electrochemical stability decreased for G3 without FEC and improved with the G3 FEC/15A FEC.

## 4. Conclusions

Binder-free HC SSEs were synthesized from cellulose-based FPs through pyrolysis and compared to HC SSEs from commercial carbon fabric. The influence of electrode thickness/macroporosity and the electrolyte additive (FEC) on the electrochemical performance and SEI layer formation was studied. All materials have similar structure, surface chemistry, and defect properties, except Kynol which has significantly lower active sites and O-surface groups. Differences between samples are found in morphology (thickness and fiber size) as well as texture (macropore volume and SSA). Kynol is significantly thicker (highest mass loading) and has the highest average fiber diameter, while GA shows a large variation in fiber diameter. However, these two materials possess the lowest specific surface area compared to G3 and 15A. Regardless of the use of the FEC additive in the $NaPF_6$ electrolyte, all materials deliver high iCE (84–93%) and RC (283–327 mAh $g^{-1}$), with the best performance achieved for HC SSE G3, which presents low macropore volume and high graphite interlayer space. However, electrode thickness and the additive strongly impacted the cycle stability. In the absence of FEC, the cycle stability decreased with the increase in electrode thickness, with thicker electrodes (>79 µm, G3, GA, and especially Kynol) facing severe fading (after 1–5 cycles), while thinner electrodes (15A) could cycle for a maximum of 50 cycles. In the presence of the FEC additive, the cycling stability is significantly improved for all electrodes (100 cycles tested), except for Kynol which is still very limited (5 cycles). This improvement in cycling behavior in the presence of FEC is due to the formation of an inorganic-rich SEI layer (instead of an organic-rich SEI layer in the absence of FEC), which is more homogeneous, thinner, stable, and conductive, ensuring cycling stability. The presence of FEC was also beneficial for rate capability, with thinner electrodes providing better performance than thicker ones, especially at high C-rate regimes. In conclusion, this work demonstrates that eco-friendly electrodes—free from binders, solvents, additives, and current collectors—can be rationally designed with controlled properties and optimized interphases to achieve high-performance and cost-effective sodium-ion battery electrodes.

## Data Availability Statement

The supplementary data for this article are attached as part of the **Supporting Information**.

## Supporting Information

The Supporting information is available free of charge at

https://pubs.acs.org/doi/10.1021/acsami.5c25017.

This includes content related to Introduction**:** State-of-art of HC SSEs; Experimental part: Properties of precursors used; manufacturing process and images of HC SSEs; detailed descriptions of Hg intrusion porosimetry, gas adsorption and TPD-MS analysis methods; Results and discussion: EDS results of HC SSEs; cross-section SEM images of G3, GA and Kynol; XRD results of G3 samples produced in different batches; deconvolution Raman spectra of HC SSEs; HR-TEM images of G3 and GA; TPD-MS desorbed gas profiles of HC SSEs; images describing the Na storage mechanism; correlation between cycle number and properties of HC SSEs; reversible capacity decay rate per cycle and symmetrical C-rate regime test of HC SSEs; XPS analysis of pristine precursors of 15A and G3; deconvolution XPS O 1s and F 1s peaks of the samples; SEM images of the HC SSEs with FEC after cycling; EDX mapping of G3 with FEC; fitted impedance spectral data with equivalent circuit; List of references (**PDF**)

## Notes

The authors declare no competing financial interest

## Acknowledgements

The authors gratefully acknowledge the financial support for the project from Région Grand-Est (project no. 23-DOC–109) and ANR (ANR–10-LABX-76-01, STORE-EX). The authors would like to express their special thanks to Dr. Jean-Marc Le Meins (XRD), Dr. Habiba Nouali (Hg intrusion porosimetry), Dr. Loic Vidal (HR-TEM & EDX) and Ms. Bénédicte Réty (TPD-MS) for their technical support through the IS2M technical platform.